\documentclass[a4paper]{article}
\usepackage{ISCSLP2026}
\usepackage{ifthen}
\newboolean{blind}
\setboolean{blind}{true} 
\title{DualSpecSE: A Dual-Path Speech Enhancement Network Integrating Mel and Complex Spectrograms}

\name{
    Xingchen Li$^1$, 
    Ziqian Wang$^1$, 
    Zikai Liu$^1$, 
    Yike Zhu$^1$, 
    Zihan Zhang$^2$, 
    Longshuai Xiao$^2$,
    Lei Xie$^1$
}
\address{
    Audio, Speech and Language Processing Group (ASLP@NPU), School of Computer Science, Northwestern Polytechnical University, China$^1$\\
   Huawei Technologies Co., Ltd., China$^2$
}

\email{
    lixingchen@mail.nwpu.edu.cn,
    lxie@nwpu.edu.cn
}

\usepackage{comment}
\usepackage{multirow}

\usepackage{cleveref}
\crefname{figure}{Fig.}{Figs.}
\Crefname{figure}{Fig.}{Figs.}
\crefname{table}{Tab.}{Tabs.}
\Crefname{table}{Tab.}{Tabs.}
\crefname{equation}{Eq.}{Eqs.}
\Crefname{equation}{Eq.}{Eqs.}
\crefname{section}{Section}{Secs.}
\Crefname{section}{Section}{Secs.}
\crefname{appendix}{Appendix}{Appendices}
\Crefname{appendix}{Appendix}{Appendices}
\begin{document}

\maketitle
\begin{abstract}
  In this paper, we propose DualSpecSE, a speech enhancement framework that jointly models Mel-spectrogram and complex spectrogram in a dual-path architecture for improved ASR performance and higher-quality speech reconstruction. The Mel branch learns coarse-grained acoustic representations and produces enhanced Mel-spectrograms for direct ASR usage, while the complex branch refines fine-grained spectral details for high-fidelity waveform reconstruction. Built upon the cross-band and narrow-band blocks from CleanMel, DualSpecSE introduces an interaction module and a fusion module to enable effective information exchange between the two branches. The model simultaneously outputs enhanced Mel and complex spectrogram without requiring a pretrained vocoder. Experimental results demonstrate consistent improvements in speech fidelity, perceptual quality, and ASR performance. Codes and audio samples are available\footnote{https://stellanli.github.io/DualSpecSE-demo/}.
\end{abstract}

\section{Introduction}

Speech enhancement aims to improve the perceptual quality of degraded speech signals. It has been widely adopted in various applications, including human-machine interaction systems, hearing assistive devices, and as a front-end module for automatic speech recognition (ASR). In recent years, significant progress has been made in speech enhancement methods based on deep learning~\cite{hu20g_interspeech, hao2021fullsubnet, lu23e_interspeech}. Most existing speech enhancement approaches operate either in the time-frequency domain~\cite{hao2021fullsubnet, 8682834, abdulatif2024cmgan, rong2024gtcrn} or directly on time-domain waveforms~\cite{pascual17_interspeech, kim21h_interspeech, kong2022speech}. These methods typically learn a direct mapping from noisy speech to clean speech representations, leading to substantial enhancement performance.

Some approaches perform speech enhancement in two stages. They first operate in the ERB or Mel domain by estimating perceptually motivated gains to enhance the spectral envelope, and then apply pitch filtering or deep filtering to further refine periodic components~\cite{valin2018hybrid, valin20_interspeech, schroter23b_interspeech}. This two-stage design reduces computational complexity by operating on a low-dimensional spectral representation. However, the limited spectral resolution of ERB- or Mel-domain gains inherently restricts the performance of these methods.

Recently, several studies have explored speech enhancement approaches that target Mel-spectrograms as the prediction objective~\cite{11097896, yang25k_interspeech}. These works have demonstrated that speech enhancement in the Mel domain is more effective at preserving coarse-grained acoustic components, thereby leading to improved ASR performance.

The Mel-spectrogram serves as a more compact speech representation compared to more complete forms such as the magnitude spectragram, complex spectragram, or waveform. Due to its compressed and perceptually motivated structure, it has been shown in several prior studies~\cite{11097896, yang25k_interspeech} to be easier to model and learn, particularly in speech enhancement and ASR-oriented scenarios. Although it performs a degree of compression on the sparse spectral representation, it preserves the semantic content of speech effectively. Under the same parameter budget, Mel-spectrograms can be directly used as inputs to ASR systems, often leading to improved recognition performance. Furthermore, when waveform reconstruction is required, a pre-trained vocoder can be employed to synthesize the corresponding speech signal.

Although these methods offer the aforementioned advantages, they still suffer from several limitations. First, they require an additional pre-trained vocoder to reconstruct the waveform. The two-stage pipeline inevitably introduces cascading errors, which degrade the naturalness of the synthesized speech. Moreover, the vocoder-generated waveform may lose the fine-grained magnitude details and phase information of the original speech, leading to reduced fidelity. In addition, the incorporation of a vocoder increases the overall model complexity and computational overhead.

In this paper, we propose DualSpecSE, a dual-path framework that jointly predicts enhanced Mel and complex spectrograms without relying on an external vocoder. The Mel branch captures ASR-friendly coarse-grained features and directly outputs enhanced Mel-spectrograms that can be fed into an ASR model for more accurate recognition, while the complex branch recovers fine-grained spectral details for high-fidelity reconstruction. An interaction module injects Mel features into the complex branch to guide detail compensation, and a fusion module integrates both representations to reconstruct the final enhanced complex spectrogram. Benefiting from the Mel branch, the enhanced speech better preserves essential speech components with less distortion. Experimental results demonstrate that, compared with single-branch models, DualSpecSE consistently improves speech quality and fidelity, while also effectively enhancing ASR performance.

\begin{figure*}[ht]
  \centering

  \includegraphics[width=0.85\textwidth]{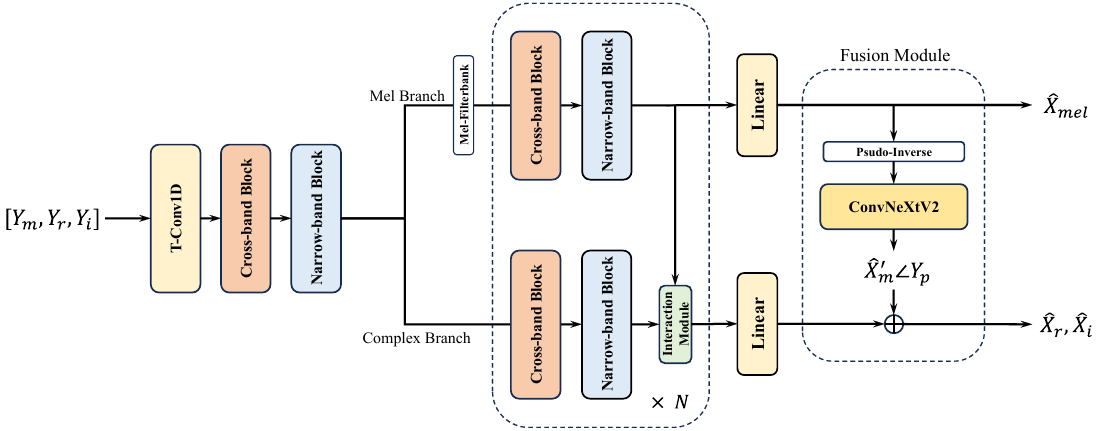}
  \caption{An overview of the dual-path architecture in DualSpecSE. }
  \label{fig:overview4}
  
\end{figure*}

\begin{figure}[ht]
  \centering

  \includegraphics[width=0.42\textwidth]{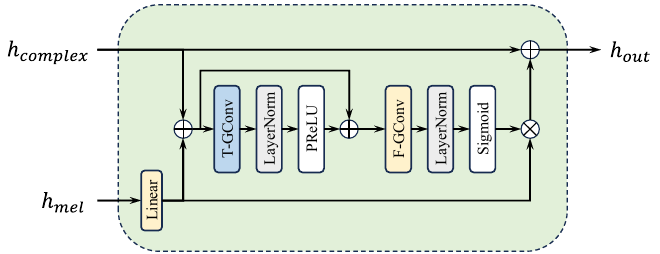}
  \caption{Interaction module. }
  \label{fig:interaction}
  
\end{figure}

\section{Method}

\subsection{Overview}

The overall architecture of DualSpecSE is illustrated in \Cref{fig:overview4}. The model consists of a shared encoder, a complex branch, and a Mel branch, which interact through the proposed interaction and fusion modules. The backbone is built upon the cross-band and narrow-band blocks from CleanMel~\cite{11097896}, with minor modifications for the dual-branch design.

Given a noisy waveform $ y \in \mathbb{R}^{L \times 1} $, we first apply STFT to obtain the complex spectrogram $ Y_o \in \mathbb{R}^{T \times F \times 2} $. Power-law compression is then applied:
\begin{equation}
    Y = \alpha \lvert Y_o \rvert ^c e^{jY_p} = Y_m e^{jY_p} = Y_r + jY_i
\end{equation}
where $Y_m, Y_p, Y_r$, and $Y_i$ denote the magnitude, phase, real part, and imaginary part, respectively. $\alpha \in \mathbb{R}^{+}$ controls the feature scale and $c \in (0,1)$ is the compression exponent. We set $ \alpha = 0.3 $ and $ c = 0.3 $ in all experiments.

The magnitude $Y_m$ and complex components $Y_r, Y_i$ are concatenated as $ Y_{in} \in \mathbb{R}^{T \times F \times 3} $ and fed into the encoder. A temporal convolution with kernel size 5 first produces an intermediate representation $ h \in \mathbb{R}^{T \times F \times H} $, which is subsequently processed by a cross-band block and a narrow-band block to model inter- and intra-band dependencies.

The encoder output is fed into the Mel branch and the complex branch in parallel to learn Mel-scale features and fine-grained linear-frequency details, respectively. The Mel branch first applies a Mel filterbank matrix $ \mathbf{W} \in \mathbb{R}^{F_{mel} \times F} $ to convert linear-frequency features into the Mel domain, reducing the frequency dimension from $F$ to $F_{mel}$. The transformed features are then processed by $N$ interleaved cross-band and narrow-band blocks to generate enhanced Mel representations. The complex branch adopts the same number of cross-band and narrow-band blocks. After each alternating block group, an interaction module incorporates features from the corresponding Mel layer, injecting Mel-based magnitude cues into the complex branch and thereby easing complex spectral modeling. 

At the output stage, each branch employs a linear projection layer to map its features to the corresponding target space. The Mel branch directly produces the enhanced Mel-spectrogram $ \hat{X}_{mel} $. In the complex branch, a fusion module further integrates Mel and complex features before projection, generating the final enhanced complex spectrogram $ \hat{X}_r, \hat{X}_i $.

\subsection{Cross-band \& Narrow-band Block}

We adopt the cross-band and narrow-band blocks proposed in CleanMel, with modifications applied to the narrow-band block.

The cross-band block models inter-frequency dependencies by processing each time frame independently along the frequency dimension with shared parameters. It consists of cascaded frequency convolution layers and across-frequency linear layers. Each frequency convolution layer includes LayerNorm, grouped 1D convolution along the frequency axis (F-GConv1d), and PReLU activation. The across-frequency linear layers first compress the channel dimension via linear projection, apply channel-wise 1D convolution, and then restore the original feature dimension.

The narrow-band block captures temporal dependencies for each frequency bin using bidirectional Mamba modules, whose outputs are averaged. To reduce computational cost, we replace the original Mamba with a GroupMamba variant that divides the feature dimension into $G$ groups, each processed independently by a separate Mamba module.

\subsection{Interaction Module}

Since Mel features constitute a compact and relatively easy-to-learn representation, whereas the complex branch is required to model more intricate complex-valued spectral characteristics, the interaction module is designed to allow the Mel branch to effectively guide the learning process of the complex branch. In this way, the interaction mechanism reduces the learning difficulty of the complex branch while encouraging it to focus on fine-grained details that may be overlooked by Mel-based representations.

Motivated by this observation, we adopt the interaction operation proposed in \cite{zhou2023novel} to map intermediate representations from the Mel branch to the complex branch, and further modify the original interaction module to accommodate the cross-scale interaction between Mel and complex spectral features, as shown in \Cref{fig:interaction}. Specifically, the Mel branch hidden representation $h_{mel}$ is first projected through a linear layer to match the feature dimensionality of the complex branch representation $h_{complex}$. The aligned features from the two branches are first added together and then fed into a temporal convolution module with a residual connection, followed by a frequency convolution module to predict an interaction mask $M$.

The temporal convolution module consists of a temporal grouped convolution layer T-GConv, a LayerNorm layer, and a PReLU activation function, while the frequency grouped convolution module is composed of a frequency convolution layer F-GConv, a LayerNorm layer, and a Sigmoid activation function. The final fused hidden representation is obtained as
\begin{equation}
    h_{\text{out}} = h_{\text{complex}} + h_{\text{mel}} \odot M 
\end{equation}
where $\odot$ denotes the element-wise multiplication.

\subsection{Fusion Module}

The Mel and complex branches operate on different spectral scales (Mel vs. linear frequency), resulting in an inherent representational gap. To bridge this mismatch, we design a fusion module that converts Mel features into the complex domain and enables joint refinement for waveform reconstruction. 

Inspired by FreeV~\cite{lv24_interspeech}, we first apply a pseudo-inverse transformation $\mathbf{W}^+$ based on the Mel filterbank matrix $\mathbf{W}$ to the enhanced Mel-spectrogram to obtain an approximated magnitude spectrogram. Since this approximation is already close to the target magnitude, a lightweight ConvNeXtV2-based~\cite{woo2023convnext} module is used to learn the residual and produce a coarse magnitude estimate $X'_m$. The estimated magnitude is then combined with the noisy phase $Y_p$ to reconstruct a complex spectrogram.

Finally, this reconstructed spectrogram is further refined by element-wise addition with the output of the complex branch. In this way, the model performs coarse magnitude correction followed by residual refinement in the complex domain, resulting in a more accurate and high-fidelity enhanced spectragram.

\subsection{Loss Functions}

We define multiple loss functions to jointly supervise the training of DualSpecSE from different spectral perspectives. For the Mel branch, the predicted Mel-spectrogram is constrained using an $ L1 $ loss on the log-Mel domain:
\begin{equation}
    X_{\text{logmel}} = \log \left( max\{X_{\text{mel}}, \epsilon\} \right)
\end{equation}
\begin{equation}
    \mathcal{L}_{\text{mel1}} = \mathbb{E}_{X_{logmel}, \hat{X}_{logmel}} \left[ \lVert X_{logmel} - \hat{X}_{logmel} \rVert_1 \right]
\end{equation}
where $\epsilon$ is a small constant for numerical stability, and is set to 1e-5 in this work. $X_{logmel}, \hat{X}_{logmel}$ denote the ground-truth and enhanced log-Mel spectrograms, respectively, which are obtained by applying a logarithmic transformation to the corresponding Mel-spectrograms.

To stabilize training, we further apply an additional magnitude loss to the refined magnitude spectrum $X_m'$ obtained from the Mel branch:
\begin{equation}
    \mathcal{L}_{\text{mag1}} = \mathbb{E}_{X_{m}, \hat{X}_{m}'} \left[ {\lVert X_{m} - \hat{X}_{m}'} \rVert_2 \right ]
\end{equation}
For the complex branch, the estimated complex spectrogram is supervised using a combination of Mel-domain loss, magnitude loss, and complex spectral loss:
\begin{flalign}
    & \mathcal{L}_{\text{mel2}} = \mathbb{E}_{X_{logmel}, \hat{X}_{logmel}'} \left[ \lVert X_{logmel} - \hat{X}_{logmel}' \rVert_1 \right] \\
    & \mathcal{L}_{\text{mag2}} = \mathbb{E}_{X_{m}, \hat{X}_{m}} \left[ {\lVert X_{m} - \hat{X}_{m}} \rVert_2 \right ] \\
    & \mathcal{L}_{\text{com}} = \mathbb{E}_{X_{r}, \hat{X}_{r}} \left[ \lVert X_{r} - \hat{X}_{r} \rVert_2 \right] + \mathbb{E}_{X_{i}, \hat{X}_{i}} \left[ \lVert X_{i} - \hat{X}_{i} \rVert_2 \right]
\end{flalign}
where $\hat{X}_{logmel}'$ denotes the log-Mel spectrogram computed from the enhanced complex spectrogram.

The overall training objective is defined as a weighted sum of all loss terms:
\begin{equation}
    \mathcal{L} = \lambda_1 \mathcal{L}_{\text{mel1}} + \lambda_2 \mathcal{L}_{\text{mag1}} + \lambda_3 \mathcal{L}_{\text{com}}  + \lambda_4 \mathcal{L}_{\text{mel2}} + \lambda_5 \mathcal{L}_{\text{mag2}}
\end{equation}
where $\lambda_1, ..., \lambda_5$ are scalar weighting factors.

\begin{table*}[t]
\centering
\caption{Comparison with other methods on the DNS Challenge 2020 test set.}
\label{tab:dns_comparison}
\resizebox{0.77\textwidth}{!}{
\begin{tabular}{l c c c c c c c c}
\toprule
\multirow{2}{*}{Method}
& \multirow{2}{*}{\#Param (M)}
& \multirow{2}{*}{FLOPs (G/s)}
& \multirow{2}{*}{WB-PESQ}
& \multirow{2}{*}{NB-PESQ}
& \multirow{2}{*}{ESTOI}
& \multicolumn{3}{c}{DNSMOS} \\
\cmidrule(lr){7-9}
& & & & & & SIG & BAK & OVRL \\
\midrule
Noisy & -- & -- & 1.58 & 2.16 & 0.809 & 3.38 & 2.31 & 2.33 \\
\midrule
FullsubNet & 14.6 & 157.9 & 2.82 & 3.39 & 0.924 & 3.53 & 4.02 & 3.24 \\
TF-GridNet & 8.52 & 300.0 & 3.12 & 3.63 & 0.935 & 3.56 & 4.14 & 3.34 \\
StoRM & 55.1 & 4600 & 2.60 & 3.17 & 0.915 & 3.57 & 4.02 & 3.30 \\
PGUSE & 5.1 & 26.3 & 3.17 & 3.67 & 0.935 & 3.55 & 4.11 & 3.33 \\
SpatialNet & 1.6 & 46.3 & 3.10 & 3.59 & 0.926 & 3.54 & 4.12 & 3.33 \\
CleanMel & 2.5+13.2 & 32.9+3.3 & 2.91 & 3.48 & 0.920 & \textbf{3.60} & 4.13 & \textbf{3.37} \\
\midrule
\textbf{DualSpecSE} & 1.85 & 30.0 & \textbf{3.25} & \textbf{3.68} & \textbf{0.936} & \underline{3.58} & \textbf{4.15} & \textbf{3.37} \\
\bottomrule
\end{tabular}
}
\end{table*}

\begin{table}[t]
\centering
\caption{Comparison with other methods on the ChiME4 test set.}
\label{tab:chime4}
\resizebox{0.40\textwidth}{!}{
\begin{tabular}{l c c c}
\toprule
Method & Output & WER (simu) (\%) & WER (real) (\%) \\
\midrule
Noisy & -- & 17.64 & 16.32 \\
\midrule
FullsubNet & waveform & 21.79 & 20.99 \\
TF-GridNet & waveform & 20.26 & 20.15 \\
StoRM & waveform & 23.00 & 22.98 \\
PGUSE & waveform & 22.31 & 23.61 \\
SpatialNet & waveform & 22.34 & 21.30 \\
CleanMel & waveform & 15.95 & 14.03 \\
\midrule
\textbf{DualSpecSE} & waveform & \textbf{14.76} & \textbf{13.21} \\
\midrule
\midrule
CleanMel & Mel-spectrogram & \textbf{13.62} & 12.21 \\
\textbf{DualSpecSE} & Mel-spectrogram & 13.92 & \textbf{12.19} \\
\bottomrule
\end{tabular}
}
\end{table}

\begin{table}[t]
\centering
\caption{Results of the ablation study on DNS Challenge 2020 test set and ChiME4 test set.}
\label{tab:ablation}
\resizebox{0.45\textwidth}{!}{
\begin{tabular}{l c c c c c}
\toprule
Method & WB-PESQ & NB-PESQ & ESTOI & OVRL & WER (simu) (\%)  \\
\midrule
DualSpecSE & \textbf{3.25} & \textbf{3.68} & \textbf{0.936} & \textbf{3.37} & \textbf{14.76} \\
\hspace{1em} w/o Mel branch & 3.12 & 3.56 & 0.929 & 3.32 & 19.05 \\
\hspace{1em} w/o Interaction module & 3.20 & 3.59 & 0.932 & 3.35 & 16.70 \\
% w/o Complex branch & -- & -- & -- & -- \\
\hspace{1em} w/o Fusion block & 3.22 & 3.62 & 0.934 & 3.33 & 17.52 \\
\bottomrule
\end{tabular}
}
\end{table}

\section{Experiments}

\subsection{Dataset}

We train DualSpecSE and the baseline models using simulated noisy–clean speech pairs generated from publicly available speech corpora, noise datasets, and room impulse responses (RIRs). The clean speech data consist of three sources: the DNS3 Challenge dataset~\cite{reddy2020interspeech} with DNSMOS scores higher than 3.4, the full EARS dataset~\cite{richter24_interspeech}, and the Emilia dataset~\cite{he2025emilia}, from which both English and Chinese utterances with DNSMOS~\cite{reddy2021dnsmos} scores above 3.6 are selected. In total, the clean speech corpus amounts to 387 hours.

The noise data are collected from the DNS3 Challenge, ESC~\cite{piczak2015dataset}, and FSD datasets~\cite{xie2024fsd}, resulting in a total duration of 376 hours. For room acoustic simulation, we use RIRs provided by SLR26 and SLR28~\cite{ko2017study}. The simulated dataset is split into training and validation sets with a ratio of 9:1, which are used for model training and validation, respectively.

For evaluation, we assess speech enhancement performance on the Interspeech 2020 DNS Challenge test set. In addition, ASR performance is evaluated on the simulated test set of the CHiME-4~\cite{vincent20164th} corpus using a pretrained  ASR model provided by ESPnet \footnote{https://github.com/espnet/espnet/tree/master/egs2}.

\subsection{Metrics}

We evaluate the quality of enhanced speech using widely adopted metrics in the speech enhancement literature, including reference-based measures such as wide-band PESQ (WB-PESQ), narrow-band PESQ (NB-PESQ)~\cite{rix2001perceptual}, and extended short-time objective intelligibility (ESTOI)~\cite{jensen2016algorithm}, as well as the non-intrusive metric DNSMOS~\cite{reddy2021dnsmos}, which reports three scores: SIG, BAK, and OVRL. In addition, we report the word error rate (WER) to assess downstream ASR performance. All metrics are reported such that higher values indicate better performance, except for WER, where lower values are better.

\subsection{Configurations}

In this work, all speech signals are sampled at 16 kHz. The short-time Fourier transform (STFT) is computed using a Hann window with a window length of 512 samples and a hop size of 128. The number of Mel frequency bins is set to $ F_{mel}=80 $, and power Mel-spectrograms are used, consistent with the configuration of the ASR model employed in our experiments.

The hidden dimension is set to $ H=64 $. The numbers of cross-band and narrow-band blocks in both the Mel branch and the complex branch are set to $ N=7 $. The number of groups in all grouped convolution layers (including T-GConv1d and F-GConv1d) is set to 8. The number of groups in GroupMamba is set to $ G=2 $. The hyper-parameters of the final loss $ \lambda_1, ..., \lambda_5 $ are set to 0.05, 1, 0.5, 0.01, 0.5. For training, we adopt the AdamW optimizer with an initial learning rate of 0.001, which is exponentially decayed according to $ lr=0.001 \times 0.99^{epoch} $. The training data are constructed using dynamic simulation. The batch size is set to 16, with 50,000 samples per epoch. The model is trained for 200 epochs.

\subsection{Comparison models}

We compare our method with several state-of-the-art speech enhancement models. FullSubNet~\cite{hao2021fullsubnet} is an LSTM-based network that integrates full-band and sub-band modeling. TF-GridNet~\cite{wang2023tf} adopts a grid-like architecture that alternately models temporal and frequency dependencies. StoRM~\cite{lemercier2023storm} is a diffusion-based stochastic regeneration model. PGUSE~\cite{zhang2025composite} combines discriminative enhancement with diffusion-based generative modeling. SpatialNet~\cite{quan2024spatialnet} interleaves narrow-band and cross-band blocks for spectral modeling. CleanMel~\cite{11097896} further adapts the SpatialNet architecture for Mel-spectrogram enhancement, we adopt the CleanMel-S-map configuration from the original paper for comparison.

\subsection{Expermental Results}

\Cref{tab:dns_comparison} reports the speech quality evaluation results on the DNS Challenge 2020 test set. Compared with the baseline CleanMel, the proposed DualSpecSE achieves significant improvements on similarity-based metrics while using fewer model parameters and lower computational complexity. Moreover, DualSpecSE outperforms SpatialNet, which is the linear-frequency counterpart of CleanMel, demonstrating the effectiveness of the proposed dual-path architecture. In addition, DualSpecSE consistently surpasses other state-of-the-art speech enhancement models, including FullSubNet, TF-GridNet, and StoRM, further validating the effectiveness of the proposed network design.

\Cref{tab:chime4} presents the ASR evaluation results on the CHiME-4 test set. We conduct evaluations on both the “simu” and “real” subsets of the test set, which correspond to the simulated and real-recorded datasets, respectively. The “Output” column indicates the type of features produced by each enhancement model. We compare CleanMel and DualSpecSE under two output settings: waveform and Mel-spectrogram. When Mel-spectrograms are used as output features, DualSpecSE achieves lower word error rate (WER) comparable to that of CleanMel, indicating that the proposed model preserves the ASR-friendly characteristics of Mel-based enhancement. When waveform outputs are considered, DualSpecSE slightly outperforms CleanMel in terms of ASR accuracy. This improvement indicates that, compared with CleanMel, DualSpecSE better preserves speech fidelity and mitigates error accumulation in cascaded enhancement pipelines.

We conduct ablation studies to evaluate the effectiveness of the proposed modules. Specifically, we remove the Mel branch, the interaction module, and the fusion module from the full DualSpecSE, respectively. The results are summarized in \Cref{tab:ablation}. The WER (simu) is evaluated on the ChiME-4 test set, while all other metrics are reported on the DNS Challenge 2020 test set. Removing the Mel branch leads to consistent performance degradation across all metrics, indicating that the coarse-grained acoustic features learned by the Mel branch are crucial for preserving rich speech components. Furthermore, eliminating either the interaction module or the fusion module also results in noticeable performance drops, demonstrating the effectiveness of both modules in enhancing the overall model performance. 

\section{Conclusions}

In this paper, we proposed DualSpecSE, a dual-path speech enhancement network that jointly models Mel and complex spectrograms to produce enhanced representations in both domains. The Mel branch captures coarse-grained, ASR-friendly features, while the complex branch refines fine-grained spectral details for high-fidelity reconstruction. Benefiting from the perceptually motivated Mel representation, DualSpecSE better preserves essential speech components during enhancement. The enhanced Mel-spectrograms can be directly used for ASR, while the enhanced complex spectrograms enable accurate waveform reconstruction.

% \section{Acknowledgements}

% The ISCA Board would like to thank the organizing committees of the past ISCSLP conferences for their help and for kindly providing the template files. \\
% Note to authors: Authors should not use logos in the acknowledgement section; rather authors should acknowledge corporations by naming them only.

\bibliographystyle{IEEEtran}

\bibliography{mybib}

% \begin{thebibliography}{9}
% \bibitem[1]{Davis80-COP}
%   S.\ B.\ Davis and P.\ Mermelstein,
%   ``Comparison of parametric representation for monosyllabic word recognition in continuously spoken sentences,''
%   \textit{IEEE Transactions on Acoustics, Speech and Signal Processing}, vol.~28, no.~4, pp.~357--366, 1980.
% \bibitem[2]{Rabiner89-ATO}
%   L.\ R.\ Rabiner,
%   ``A tutorial on hidden Markov models and selected applications in speech recognition,''
%   \textit{Proceedings of the IEEE}, vol.~77, no.~2, pp.~257-286, 1989.
% \bibitem[3]{Hastie09-TEO}
%   T.\ Hastie, R.\ Tibshirani, and J.\ Friedman,
%   \textit{The Elements of Statistical Learning -- Data Mining, Inference, and Prediction}.
%   New York: Springer, 2009.
% \bibitem[4]{YourName17-XXX}
%   F.\ Lastname1, F.\ Lastname2, and F.\ Lastname3,
%   ``Title of your ISCSLP 2026 publication,''
%   in \textit{ISCSLP 2026 -- 23\textsuperscript{rd} Annual Conference of the International Speech Communication Association, September 18-22, Incheon, Korea, Proceedings, Proceedings}, 2026, pp.~100--104.
% \end{thebibliography}

\end{document}